\documentclass{svjour2}                    % onecolumn
\smartqed  % flush right qed marks, e.g. at end of proof
\usepackage{color}
\usepackage{graphicx}
\usepackage{mathtools}
\usepackage{amsmath}
\usepackage{amssymb}
\usepackage{array}
\usepackage{bbm}
\usepackage{subcaption}
\usepackage{enumitem}

\usepackage{dsfont} % for unit operator  (indicator function) \mathds{1}

\usepackage{hyperref}

\def\xstrut{\rule[-2ex]{0pt}{5ex}}

\newcommand{\be}{\begin{equation}}
\newcommand{\ee}{\end{equation}}
\newcommand{\bel}[1]{\begin{equation}\label{#1}}
\newcommand{\DUMMY}[1]{\end{equation}\label{#1}}

\newcommand{\bea}{\begin{eqnarray}}
\newcommand{\eea}{\end{eqnarray}}
\newcommand{\balign}{\begin{align}}
\newcommand{\ealign}{\end{align}}
\newcommand{\ba}{\begin{array}}
	\newcommand{\ea}{\end{array}}
\newcommand{\bfig}{\begin{figure}}
	\newcommand{\efig}{\end{figure}}

\allowdisplaybreaks

\newcommand{\iid}{{\it i.i.d. }}
\newcommand{\rmd}{\mathrm{d}}

\newcommand{\Z}{{\mathbb Z}}

\newcommand{\abs}[1]{\left|{{#1}}\right|}

\newcommand{\diag}[1]{\text{diag}\left({{#1}}\right)}

\newcommand{\expct}[1]{\left\langle{{#1}}\right\rangle}
\newcommand{\occ}[2]{n_{{#1}}({{#2}})}
\newcommand{\occR}[3]{n^{({{#1}})}_{{#2}}({{#3}})}

\newcommand{\lattice}{\Gamma}
\newcommand{\length}{L}
\newcommand{\pDir}{\probT}
\newcommand{\diffMat}{\mathrm{D}}

\newcommand{\diffTuniv}{\overline{D}_\TA}
\newcommand{\diffTp}[1]{D_{\TA,{#1}}}
\newcommand{\diffSuniv}{\overline{D}_\SE}
\newcommand{\LogScale}{E}
\newcommand{\LogScaleuniv}{\overline{E}}
\newcommand{\LogScalep}[1]{\LogScale_{{#1}}}

\newcommand{\diffSp}[1]{D_{\SE,{#1}}}

\newcommand{\density}{\rho}
\newcommand{\denFluc}[2]{u({{#1}},{{#2}})}
\newcommand{\secClDensity}[1]{\rho^\mathrm{2nd~cl.}\left({#1}\right)}
\newcommand{\curr}{\vec{j}}

\newcommand{\corr}[2]{S({{#1}},{{#2}})}
\newcommand{\corrF}[2]{\hat{S}({{#1}},{{#2}})}
\newcommand{\estCorr}[3]{S^{{#1}}({{#3}},{{#2}})}

\newcommand{\INDEX}{I}
\newcommand{\indexa}{k}
\newcommand{\indexb}{j}
\newcommand{\indexc}{r}
\newcommand{\sitea}{\vec{0}}
\newcommand{\siteb}{\vec{x}}
\newcommand{\sitec}{\vec{y}}
\newcommand{\timea}{0}
\newcommand{\timeb}{t}
\newcommand{\timec}{t'}
\newcommand{\probT}{p}
\newcommand{\probTb}{q}
\newcommand{\varS}[2]{\sigma^2_{\SE,{{#2}}}({{#1}})}
\newcommand{\varT}[2]{\sigma^2_{\TA,{{#2}}}({{#1}})}

\newcommand{\TA}{\mathrm{T}}
\newcommand{\SE}{\mathrm{S}}

\begin{document}

\title{Logarithmic superdiffusion in two dimensional driven lattice gases \thanks{Dedicated to Herbert Spohn on the occasion of his 70th birthday.}}


\titlerunning{Logarithmic superdiffusion in two dimensional driven lattice gases}        % if too long for running head

\author{J. Krug$^{(1)}$ \and R. A. Neiss$^{(2)}$ \and \\ A. Schadschneider$^{(1)}$ \and J. Schmidt$^{(1)}$}


\institute{$^{(1)}$Institut f\"{u}r Theoretische Physik, Universit\"{a}t zu K\"{o}ln, Z\"ulpicher Str. 77,\\
D-50937 Cologne, Germany.       \\
$^{(2)}$Mathematisches Institut der Universit\"{a}t zu K\"{o}ln, Weyertal 86-90, \\
D-50931 Cologne, Germany.
}

\date{Received: date / Accepted: date / recent editing: \today}

\maketitle

\begin{abstract}
The spreading of density fluctuations in two-dimensional driven diffusive systems is marginally anomalous. Mode coupling theory predicts that the diffusivity in the direction of the drive 
diverges with time as $(\ln t)^{2/3}$ with a prefactor depending on the macroscopic current-density relation and the diffusion tensor of the fluctuating hydrodynamic field equation.
Here we present the first numerical verification of this behavior for a particular version of the two-dimensional asymmetric exclusion process. Particles jump strictly asymmetrically along
one of the lattice directions and symmetrically along the other, and an anisotropy parameter $p$ governs the ratio between the two rates. Using a novel massively parallel coupling algorithm
that strongly reduces the fluctuations in the numerical estimate of the two-point correlation function, we are able to accurately determine the exponent of the logarithmic correction.  In addition, the variation
of the prefactor with $p$ provides a stringent test of mode coupling theory.

% \PACS{05.60.Cd \and 05.20.Jj \and 05.70.Ln \and 47.10.-g}
\end{abstract}

\newpage

% #########################################################
% ####################	INTRODUCTION	###################
% #########################################################
\section{Introduction}

Low-dimensional systems with one or several conserved fields often display anomalous dynamics, in the
sense that fluctuations spread faster than diffusively. A prominent example is the phenomenon of 
long-time tails in classical fluids at thermal equilibrium \cite{Pomeau1975,Forster1977,Spohn1991}. In $d$ dimensions the velocity autocorrelation function
decays as $t^{-d/2}$ to leading order, which implies diverging transport coefficients in dimensions
$d \leq 2$. Specifically, in $d=1$ hydrodynamic modes are governed by the superdiffusive dynamic
exponent $z=3/2<2$ \cite{vanBeijeren2012,Popkov2016,Spohn2016},
whereas in the marginal case $d=2$ the corrections to normal diffusion
are only logarithmic. Mode coupling theory predicts that the diffusivity diverges as 
$(\ln t)^{1/2}$ in two dimensions, a behavior that has been notoriously difficult to verify
numerically \cite{vanderHoef1991,Lowe1995,Isobe2008}

In 1985, van Beijeren, Kutner and Spohn (BKS) discovered a similar scenario for driven diffusive
systems (DDS) characterized by a single conserved density onto which a steady current is imposed
by an external drive \cite{Beij85}. In one dimension such systems are governed by the noisy Burgers equation,
which had been studied earlier by Forster, Nelson and Stephen in the
hydrodynamic context \cite{Forster1977}
and was subsequently introduced by Kardar, Parisi and Zhang (KPZ) as a description of stochastic
interface dynamics \cite{KPZ1986}. The anomalous dynamic exponent $z=3/2$ is now recognized to be the hallmark
of the one-dimensional KPZ universality class, and the recent progress in the understanding
of one-dimensional fluids relies crucially on previous developments pertaining to the KPZ equation
and its various representatives \cite{Corwin2012,KriKru2010}. 
% \cite{vanBeijeren2012,Spohn2016}. 

However, in dimensions $d \geq 2$ the DDS and Burgers/KPZ problems are fundamentally different,
because the Burgers equation describes the isotropic evolution of a vector field (the height
gradient of the KPZ interface) whereas the DDS evolution equation is scalar and anisotropic due
to the drive. As a consequence, the strong coupling behavior that characterizes the KPZ
equation in dimensions $d \geq 2$ is absent in the DDS case \cite{Janssen1986}. Similar to classical fluids, driven 
diffusive systems display normal diffusive behavior in $d>2$ and are marginally superdiffusive in 
$d=2$. 

Based on mode coupling theory, BKS predicted that the variance of the
two-point correlation function 
in two dimensions grows as $t (\ln t)^\zeta$ with $\zeta = 2/3$. A rigorous proof of
this asymptotics was presented by Yau for the
asymmetric simple exclusion process (ASEP) at density $\rho = 1/2$ \cite{Landim2004,Yau04}. On
physical grounds one expects the same behavior to apply throughout the 
class of two-dimensional DDS, but extending the result of Yau to a
more general setting has so far remained elusive. In recent work the 
existence of logarithmic superdiffusivity has been established
for a fairly broad class of models, but only upper and lower bounds
$1/2 \leq \zeta \leq 1$ were obtained for the exponent of the logarithmic correction
\cite{Quastel2013}. 

In this situation it is of interest to explore to what extent
logarithmic superdiffusion in 
two-dimensional DDS can be ascertained using numerical simulations,
and, in particular, whether accurate estimates of the exponent $\zeta$
can be obtained. Here we report on large-scale, high-precision
simulations of the two-dimensional ASEP that achieve this goal by making use of a novel
algorithm based on coupling simulation runs starting from different
initial configurations. We believe that our methodology could be
useful also in other systems that display marginal superdiffusion, such as two-dimensional hydrodynamic lattice gas
models \cite{Landim2005} and one-dimensional DDS at densities
corresponding to an inflection point of the current-density relation \cite{Devillard1992,Binder1994}.
In both of these cases the exponent $\zeta$ is predicted to take the
value $\zeta = 1/2$. 

In the next section we will define the model used in our study,
explain the algorithm and recall the mode coupling theory of BKS. The
simulation results are presented in Sect.~\ref{Sec:Results} and some conclusions
and consequences for future work are discussed in
Sec.~\ref{Sec:Discussion}.

% #########################################################
% ####################		METHOD		###################
% #########################################################
\section{Model and methods}

\subsection{Two-dimensional asymmetric simple exclusion process}

Simulations were carried out on a two-dimensional lattice of
rectangular shape and with periodic boundary conditions. The lattice
is populated with a single species subject to the exclusion
principle. For convenience, the overall density was chosen to be
$\density=\frac12$, which ensures that the density fluctuations 
have no systematic drift. As a consequence the two-point correlation
function is symmetric and has a stationary peak. 
The time evolution is computed following a Markov chain Monte Carlo algorithm of random sequential updates.
The stochastic dynamics of the system proceeds as follows. At first,
the initial state is drawn from the equidistribution on the full
configuration space. 
This amounts to populating the sites of the
lattice according to a product measure, with no correlations apart
from finite size corrections, which moreover corresponds to the
invariant distribution of the dynamics. 
In a single random
sequential update, one independently selects a lattice site from
equidistribution and also a direction of motion. 

In our implementation
of the two-dimensional ASEP particles move strictly asymmetrically
along one lattice direction, referred to as the TASEP
direction $x_T$ in the following, and symmetrically along the perpendicular 
SSEP direction $x_S$. Here TASEP and SSEP stand for the
totally asymmetric simple exclusion process and the symmetric
simple exclusion process, respectively. 
Correspondingly, the dimensions of the rectangular
lattice $\lattice=\Z/\length_\TA\times\Z/\length_\SE$ are denoted by
$\length_\TA$ and $\length_\SE$. The parameter $\pDir\in(0,1)$ is the probability for choosing
the TASEP direction, and $(1-\pDir)/2$ is the probability for the two
directions along the SSEP axis. If the target lattice site is empty,
the particle moves, otherwise the exclusion principle prohibits the
motion and the particle remains on site. Time is counted in units of
full system sweeps, where one sweep consists of as many random sequential updates as there
are lattice sites. For $p=\frac{1}{2}$ this update rule is the finite
size equivalent to the system studied in \cite{Yau04}. Below we will
make systematic use of the control parameter $\pDir$ for extracting
the superdiffusive behavior and for a detailed comparison with the predictions of mode coupling theory.

\subsection{Coupling estimator}

The quantity of interest is the normalized two-point correlation function
\begin{equation}
\label{eqn:two-point-correlation}
\corr{\siteb}{\timeb} = \kappa^{-1}\left(\expct{\occ{\siteb}{\timeb} \occ{\sitea}{\timea}} - \density^2\right),
\end{equation}
where $\expct\cdot$ is the expectation value,
$\occ{\siteb}{\timeb}\in\{0,1\}$ is the occupation number at site
$\siteb=(x_\TA,x_\SE)^T$ at time $\timeb$, and
$\kappa=\density(1-\density)$ denotes the compressibility. 
In standard simulation approaches this two-point correlator is
estimated by averaging over all lattice sites (exploiting translation
invariance), time evolution (exploiting ergodicity), and many
independent realizations (initial states and time
evolutions). However, for $\corr{\siteb}{\timeb}$ these methods 
fail to reach statistical error bars less than $10^{-6}$
at reasonable computational cost. We estimate an upper
bound for numerically accessible system times $\timeb_{\max}$ by
requiring data for our analysis with a maximal relative error of
10\%. The correlation function can be approximated by a 2D SSEP
process whose maximum is analytically known and decays $\propto
\timeb^{-1}$. From this, we deduce that the maximally observable time
is $\timeb_{\max} \approx 5000$. 
On the other hand, logarithmic corrections of the kind that we are
concerned with here generally require longer times to become visible in the data.

This calls for new ideas to estimate the two-point correlator with
significantly lower variance. It turns out that, in this particular
case of a single species system, \textit{coupling} multiple
computational runs offers huge advantages. Coupling is a
well-established tool in the mathematical analysis of interacting
particle systems \cite{Liggett1985,Liggett1999}, but so far it does
not seem to have been exploited much for computational
studies (see however \cite{Schmidt2015} for a related example). 
Consider a set of independent initializations 
$\{\occR{\indexa}{\siteb}{\timeb=0}\}_{\indexa\in\INDEX}$ indexed by
$\INDEX$ that are coupled in the sense that for their time evolution,
in any random sequential update, the same lattice sites and the same
directions of motion are drawn in every system. In this way one random
number evolves many systems \textit{simultaneously} and the law of
motion can be expressed in bitwise AND and OR operators. This allows
one to simulate the coupled systems at no additional cost for a number
$\vert \INDEX \vert$ up to the bitwidth of the computational architecture, which is here
always equal to 64. 

\begin{figure}[h]
	\begin{subfigure}[t]{0.48\textwidth}
		\includegraphics[width=\textwidth]{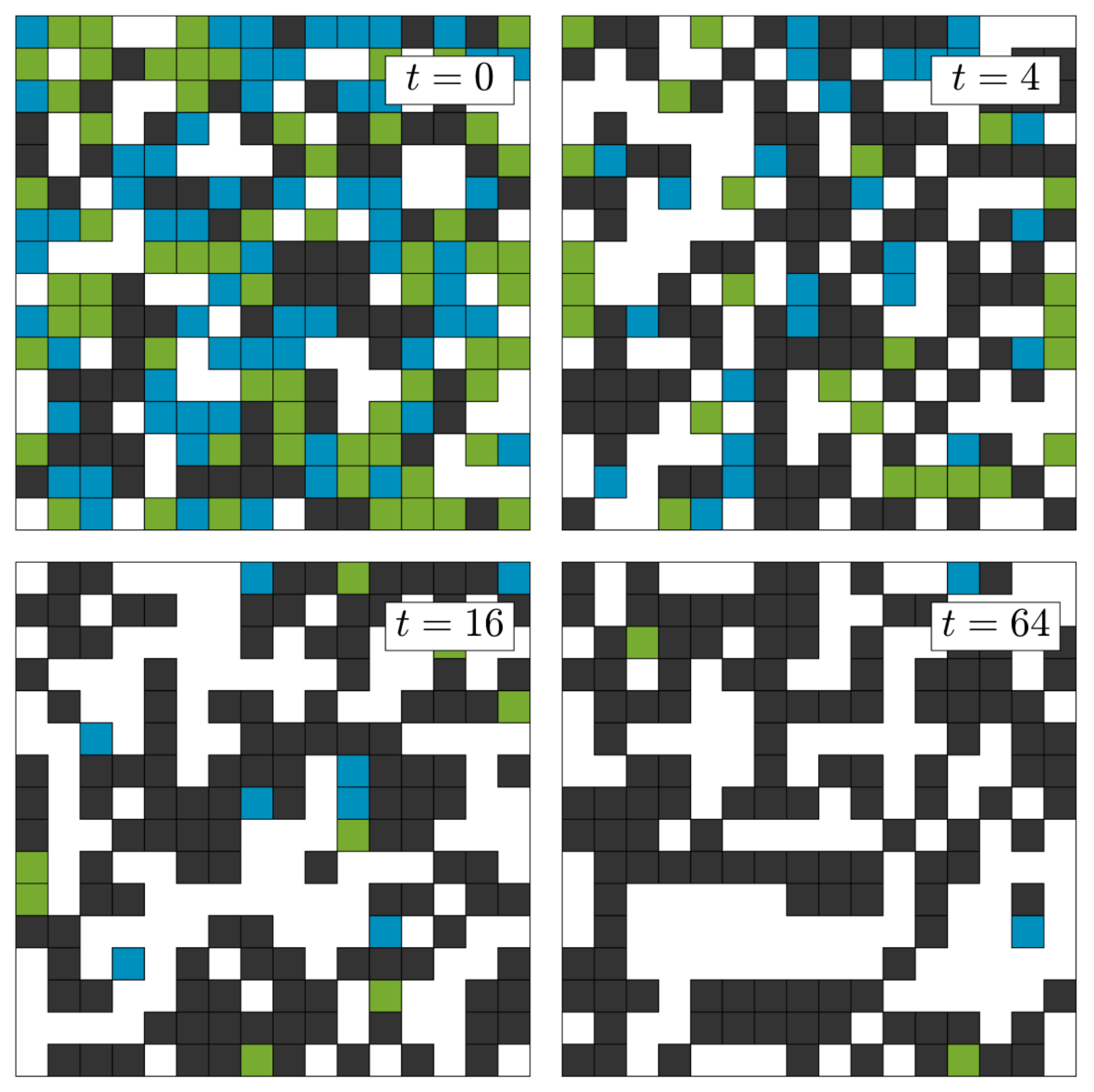}
		\centering
		\caption{}
		\label{fig:exemplary_2nd_class_evolution}
	\end{subfigure}
	\begin{subfigure}[t]{0.48\textwidth}
		\includegraphics[width=\textwidth]{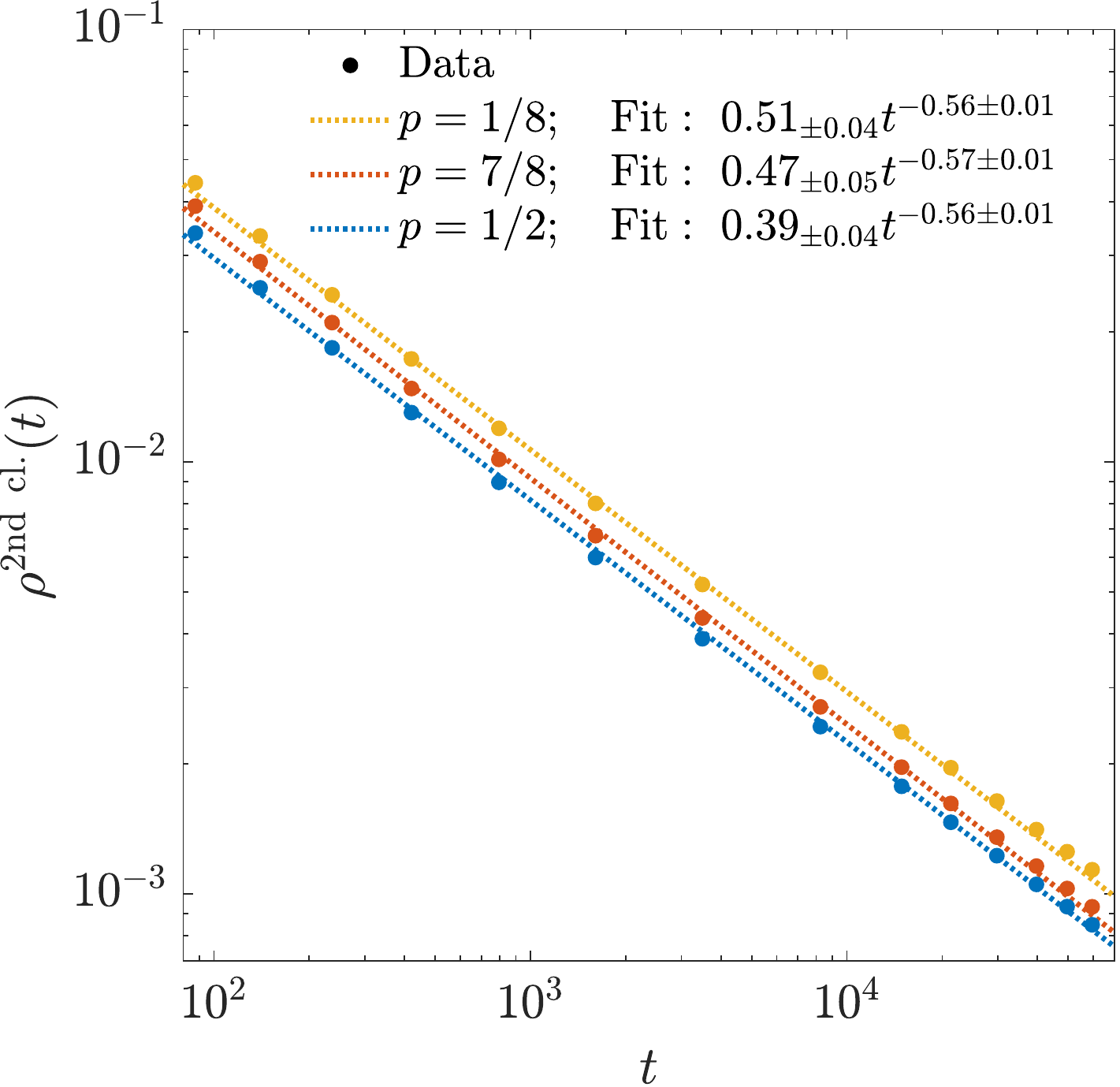}
		\centering
		\caption{}
		\label{fig:2nd_class_decay}
	\end{subfigure}
	\caption{(a) Exemplary evolution of two coupled $16\times16$ systems
		at different times $t$ showing the decay of second-class (colored)
		particles. Green and blue squares indicate that
		$\occR{1}{\siteb}{\timeb}-\occR{2}{\siteb}{\timeb}=\pm1$. As the
		overall particle number in both systems is equal and conserved, the
		two types of second class particles are equally abundant.
		(b) Decay of the global second class particle density on double
		logarithmic scales. While the decay law (\ref{Eq:DensityDecay}) seems universal,
		the prefactor $C$ is not monotonic in $\probT$ and must have a minimum in
		$(\frac18,\frac78)$. Data errors are within the symbol size,
		computed from 350 independent simulation runs.}
\end{figure}

The defining property of \textit{attractive} interacting particle systems such
as the ASEP is that the configurations of coupled systems approach
each other over time \cite{Liggett1985,Liggett1999}. It is helpful to visualize this process by
representing the discrepancies between two configurations by second
class particles.   
For two systems $\indexa,\indexb\in\INDEX$, we say there is a
discrepancy at site $\siteb$ and time $\timeb$ when
$\occR{\indexa}{\siteb}{\timeb}-\occR{\indexb}{\siteb}{\timeb}=\pm1$,
where the two possible signs define two species of second class
particles. Through the dynamics of the coupled systems, second class
particles cannot be generated. They are subject to exclusion with
first class particles defined by the condition
$\occR{\indexa}{\siteb}{\timeb}=\occR{\indexb}{\siteb}{\timeb}=1$ and
with second class particles of the same sign, and annihilate pairwise
whenever a +1 particle meets a $-$1 particle. Thus the number of
discrepancies decreases over time and approaches zero for $t \to
\infty$. In Fig.~\ref{fig:exemplary_2nd_class_evolution} we show the time
evolution of two coupled configurations illustrating this process, and
Fig.~\ref{fig:2nd_class_decay} depicts the power law decay of the overall second class particle density
\be
\label{Eq:DensityDecay}
\secClDensity{\timeb} \equiv \expct{\frac{1}{\abs{\lattice}}\sum_{\siteb\in\lattice} \abs{\occR{\indexa}{\siteb}{t}-\occR{\indexb}{\siteb}{t}}} \stackrel{t~\text{large}}{\simeq} Ct^{-\beta}, \beta=0.56_{\pm0.01}.
\ee
We are not aware of any analytic prediction for the observed exponent $\beta$.

Importantly, introducing the coupled systems allows us to express the two-point
correlation function in a different manner as
\begin{align}
\label{Eq:CorrTrick}
\kappa\corr{\siteb}{\timeb} =& \expct{\occR{\indexa}{\siteb}{\timeb} \occR{\indexa}{\sitea}{\timea}} - \density^2 
= \expct{\occR{\indexa}{\siteb}{\timeb} \occR{\indexa}{\sitea}{\timea}} - \expct{\occR{\indexb}{\siteb}{\timeb} \occR{\indexa}{\sitea}{\timea}} \nonumber \\
=& \expct{\left(\occR{\indexa}{\siteb}{\timeb} -
	\occR{\indexb}{\siteb}{\timeb}\right)
	\occR{\indexa}{\sitea}{\timea}} =
\expct{\left(\occR{\indexa}{\sitea}{\timeb} -
	\occR{\indexb}{\sitea}{\timeb}\right)
	\occR{\indexa}{-\siteb}{\timea}}, 
\end{align}
where we used that $\occR{\indexa}{\siteb}{\timea}$ and
$\occR{\indexb}{\sitec}{\timeb}$ are independent for
$\indexa\neq\indexb$, and that the system is translational
invariant. By construction, because
$\vert \occR{\indexa}{\siteb}{\timeb} -
\occR{\indexb}{\siteb}{\timeb} \vert$ is often zero for large times, this
expression can be efficiently evaluated in simulation data and 
% gives deviation 
has fluctuations \textit{declining} in time. In addition, the relation
(\ref{Eq:CorrTrick}) can be exploited for any pair of coupled systems,
giving our estimator for the coupled realizations indexed by $\INDEX$
as
\begin{align}
\label{single-estimator}
\estCorr{\INDEX}{\timeb}{\siteb} =& 
\frac12 \kappa^{-1} \binom{\abs{\INDEX}}{2}^{-1} 
\abs{\Gamma}^{-1} \times \nonumber \\
&\times \sum_{\substack{\indexa,\indexb\in\INDEX, \\ \indexa<\indexb}}  \sum_{\sitec\in\Gamma} \left(\occR{\indexa}{\sitec}{\timeb} - \occR{\indexb}{\sitec}{\timeb}\right) \left(\occR{\indexa}{\sitec-\siteb}{\timea} - \occR{\indexb}{\sitec-\siteb}{\timea}\right).
\end{align}
Because of complicated cross-correlations in this expression, we
cannot immediately estimate its variance. Nevertheless, if we have $R$
independent realizations of coupled systems indexed by
$\abs{\INDEX_1}=\abs{\INDEX_2}=\dots=\abs{\INDEX_R}$, their estimators
$\{\estCorr{\INDEX_{\indexc}}{\siteb}{\timeb}\}_{\indexc}$ are \iid
and we may use the standard mean value and error estimates.

\subsection{Mode coupling theory}
\label{Sec:MCT}
Mode coupling provides a promising inroad for an analytic study of
the two-point correlation function, for which the theoretical groundwork
was carried out by BKS \cite{Beij85}. Therein, the density fluctuations
$\denFluc{\siteb}{\timeb}$ are described using fluctuating
hydrodynamics and the normalized two-point correlation function is
expressed through the fluctuation fields as
\bel{structure_function_NLFH}
\corr{\siteb}{\timeb}=\kappa^{-1}\expct{\denFluc{\siteb}{\timeb}\denFluc{\sitea}{\timea}}.
\ee
The starting point is the continuity equation
\bel{eq:density-fluctuation}
\partial_\timeb \denFluc{\siteb}{\timeb} = \nabla\cdot\left(\diffMat
\cdot \nabla\denFluc{\siteb}{\timeb} - {\partial_\rho \curr}~ \denFluc{\siteb}{\timeb} - \frac{1}{2}~{\partial_\rho^2\curr}~\denFluc{\siteb}{\timeb}^2+B\xi\left(\siteb,\timeb\right)\right),
\ee
where the current is replaced by the steady state current-density
relation $\curr(\rho+\denFluc{\siteb}{\timeb})$ expanded up to second
order in the fluctuation. Diffusion and white noise  with correlations
$\expct{\xi\left(\siteb,\timeb\right)\xi\left(\sitec,\timec\right)}=\delta\left(\siteb-\sitec\right)\delta\left(\timeb-\timec\right)$
are added phenomenologically to account for the randomness in the time
evolution. With the continuity equation at hand, the mode coupling
formalism translates the evolution of the fluctuation fields into an
integrodifferential equation for the two-point correlation function
known as the mode coupling equation, 
\begin{align}
\label{eq:Mode_Coupling_Position}
\partial_\timeb\corr{\siteb}{\timeb}=-{\partial_\rho \curr}
\cdot\nabla \corr{\siteb}{\timeb}+\left(\nabla\cdot \diffMat \cdot
\nabla\right)\corr{\siteb}{\timeb} \nonumber \\ + \intop_0^\timeb\rmd s \intop\rmd y^\rmd \corr{\siteb-\vec{y}}{\timeb-s}  M\left( \vec{y},s\right),
\end{align}
where the memory kernel $M$ accounts for nonlinear interactions and noise.
The simplest analytically solvable approximation is the one-loop  kernel
\bel{One-Loop kernel}
M\left( \vec{y},s\right)=\frac{\kappa}{2} ( \partial_\rho^2\curr\cdot \nabla)^2 \left(\corr{\vec{y}}{s}\right)^2.
\ee
Solving the mode coupling equation for $d\geq3$ predicts nonlinear
contributions to be irrelevant and therefore a diffusive decay
as $\max_{\siteb} \corr{\siteb}{\timeb} \sim t^{-d/z}$
with the diffusive dynamical exponent $z=2$. In contrast, for $d=1$
nonlinear contributions dominate and the asymptotic behavior becomes
superdiffusive matching the exact KPZ dynamical exponent $z=3/2$ but
failing to recover the exact scaling function \cite{Prahofer2004}. Dimension
$d=2$ turns out to be the borderline dimension where the diffusion is
logarithmically enhanced in the direction of the nonlinearity
specified by the unit vector $\vec{e}_{\curr^{\prime\prime}}
= \partial_\rho^2\curr/\vert \partial_\rho^2\curr \vert$ \cite{{Beij85}}.

As controlling diffusion anisotropy will turn out to be crucial for
making the logarithmically enhanced diffusion numerically accessible, we
present a detailed solution of the two-dimensional mode coupling
equation for an arbitrary diffusion tensor. 
We first perform a Galilean transformation $\vec{\siteb}\mapsto \vec{\siteb} -t{\partial_\rho \curr}$
to remove the drift term and then transform into Fourier space. Using the convention
\bel{eq:fourier transform}
f(\vec{x})=\frac{1}{\left(2\pi\right)^{\rmd/2}}\intop \rmd k^\rmd e^{i\vec{k}\cdot\vec{x} } \hat{f}(\vec{k})
,\quad
\hat{f}(\vec{k})=\frac{1}{\left(2\pi\right)^{\rmd/2}}\intop \rmd x^\rmd e^{-i\vec{k}\cdot\vec{x} } f(\vec{x}).
\ee
the mode coupling equation becomes
\bel{eq:Mode_Coupling_Momentum}
-\partial_\timeb\corrF{\vec{k}}{\timeb}-\left(\vec{k}\cdot \diffMat \cdot \vec{k}\right)\corrF{\vec{k}}{\timeb}=  \intop_0^\timeb\rmd s \corrF{\vec{k}}{\timeb-s}  \widehat{M}\left( \vec{k},s\right),
\ee
with the memory kernel
\bel{eq:One-Loop kernel Fourier}
\widehat{M}\left( \vec{k},s\right)=\frac{\kappa\left( \partial^2_\rho \curr \right)^2}{2}\left(\vec{\mathrm{e}}_{\curr^{\prime\prime}}\cdot \vec{k} \right)^2 \intop \rmd q^\rmd \corrF{\vec{q}}{s}\corrF{\vec{k}-\vec{q}}{s}.
\ee
Similar to \cite{Beij85} we account for the logarithmically enhanced
diffusion through the scaling ansatz 
\bel{eq:F_Scaling_Ansatz}
\hat{S}(\vec{k},t)\simeq\frac{1}{2\pi}\exp\left(-(\vec{k}\cdot\diffMat\cdot\vec{k})t - E \left(\vec{\mathrm{e}}_{\curr^{\prime\prime}}\cdot \vec{k} \right)^2 t (\ln t)^{\zeta} \right).
\ee
To capture the asymptotic behavior we take the limit
$\abs{\vec{k}}\rightarrow 0$ and assume $t$ to be large in the sense
of
\bel{eq:large_t_quantification}
\ln t
\gg  \left(\frac{\det(\diffMat)}{\left( \vec{\mathrm{e}}_{\curr^{\prime\prime}} \cdot \mathrm{adj}(\diffMat)\cdot \vec{\mathrm{e}}_{\curr^{\prime\prime}}\right)E}\right)^{1/\zeta} .
\ee
Finally, evaluating the left and right hand sides of Eq.~(\ref{eq:Mode_Coupling_Momentum}) gives
\bel{eq:One_Loop_evaluation}
\frac{E \left(\vec{\mathrm{e}}_{\curr^{\prime\prime}}\cdot \vec{k} \right)^2 }{2\pi} \cdot (\ln t)^{\zeta}
\overset{!}{=}
\frac{\kappa \left(\vec{\mathrm{e}}_{\curr^{\prime\prime}}\cdot \vec{k} \right)^2}{32\pi^2\left(1-\frac{\zeta}{2}\right)\sqrt{\left( \vec{\mathrm{e}}_{\curr^{\prime\prime}} \cdot \mathrm{adj}(\diffMat)\cdot \vec{\mathrm{e}}_{\curr^{\prime\prime}}\right)E}}\cdot (\ln t)^{1-\zeta/2}
\ee
from which one deduces
\bel{eq:zeta_one_loop}
\zeta=\frac{2}{3}
\ee
and
\bel{eq:sclae_parameter_log_diffision}
\LogScale=\left( \frac{3 \kappa \left( \partial^2_\rho \curr \right)^2}{32\pi\sqrt{\vec{\mathrm{e}}_{\curr^{\prime\prime}} \cdot \mathrm{adj}(\diffMat)\cdot \vec{\mathrm{e}}_{\curr^{\prime\prime}}}} \right)^{2/3}.
\ee

Applying this approximation to the system of interest with anisotropy
parameter $\probT$, we have the current-density relation
\be
\label{Eq:CurrDens}
\curr_\probT(\rho) = \probT\density(1-\density)(1,0)^T
\ee 
and the
diffusion matrix
$\diffMat_{\probT}=\diag{\diffTp\probT,\diffSp\probT}$. Under the
reasonable assumption that the diffusion constants scale linearly with the probability of motion in the respective direction, we expect
\be
\diffSp{\probT} \equiv \diffSuniv~(1-\probT) = \frac{1-\probT}{2},~~\diffTp{\probT} = \diffTuniv~\probT
\ee
with universal constants $\diffSuniv,\diffTuniv$. As will be verified
numerically below, $\diffSuniv$ coincides with the exact diffusion
coefficient $\diffSuniv = \frac{1}{2}$ of the SSEP. With this, the
expression (\ref{eq:sclae_parameter_log_diffision}) reduces to 
\bel{eq:sclae_parameter_log_diffision_2}
\LogScalep{\probT} = \LogScale=\frac{1}{8}\left( \frac{3 p^2}{\pi\sqrt{1-p}} \right)^{2/3}
\equiv \LogScaleuniv~ \probT\left(\frac{\probT}{1-\probT}\right)^{\frac{1}{3}}
\ee
with a universal constant $\LogScaleuniv$.
Note that according to the condition of
Eq.~(\ref{eq:large_t_quantification}), the superdiffusive correction
is most easily accessible when $\LogScale$ is large, i.e. 
for $\probT \to 1$, whereas for small $p$ significant finite-time corrections are expected.

% #########################################################
% ####################		RESULTS		###################
% #########################################################
\section{Results}
\label{Sec:Results}

\subsection{Simulated systems}

Simulation data were obtained from six different settings of the
anisotropy parameter $\probT$ and the lattice dimensions (Table \ref{Table}).
They were chosen consistently such that at $t=60000$ the correlations have not
yet spread to the boundaries of the system. For this, we chose
$\abs{\lattice}\geq 3\cdot 10^7$ and the relation between the lattice
periodicities was taken to scale with the corresponding probabilities,
i.e., 
% $\lattice=\Z/\length_\TA\times\Z/\length_\SE$ with 
$\length_\TA:\length_\SE\approx\probT:1-\probT$.
The two-point correlation was recorded on a cross of lattice points through
the origin parallel to the axes, whose extension scales with the peak width.

\begin{table}
	\begin{center}
		\begin{tabular*}{\textwidth}{l >{\xstrut$}c<{$} | >{\xstrut$}c<{$} >{\xstrut$}c<{$} >{\xstrut$}c<{$} >{\xstrut$}c<{$} >{\xstrut$}c<{$} >{\xstrut$}c<{$}}
			\hline
			TASEP prob. & \probT & \frac12 & \frac14 & \frac34 & \frac18 & \frac78 & \frac{31}{32} \\ 
			\hline
			TASEP dim. & \length_\TA & 6,144 & 4,096 & 12,288 & 2,048 & 16,384 & 32,768 \\
			SSEP dim. & \length_\SE & 6,144 & 12,288 & 4,096 & 16,384 & 2,048 & 1,024 \\
			\hline
			\iid realizations & R & 1,000 & 1,000 & 1,000 & 1,000 & 1,000 & 1,000 \\
			\hline
		\end{tabular*}
	\end{center}
	\caption{\label{Table}Summary of ASEP system parameters used in this work.}
\end{table}

\subsection{Scaling function}

\begin{figure}[h]
	\centering
	\begin{subfigure}[t]{0.48\textwidth}
		\includegraphics[width=\textwidth]{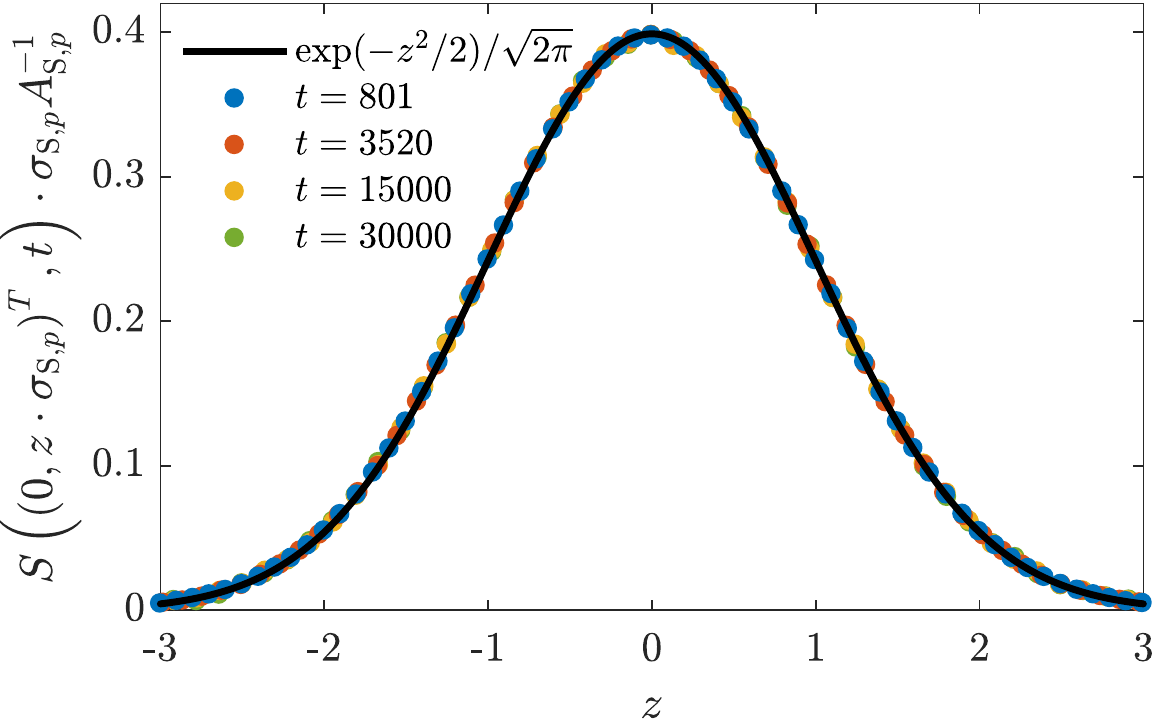}
		\caption{}
		\label{fig:1_1_sep_collaps}
	\end{subfigure}
	\begin{subfigure}[t]{0.48\textwidth}
		\includegraphics[width=\textwidth]{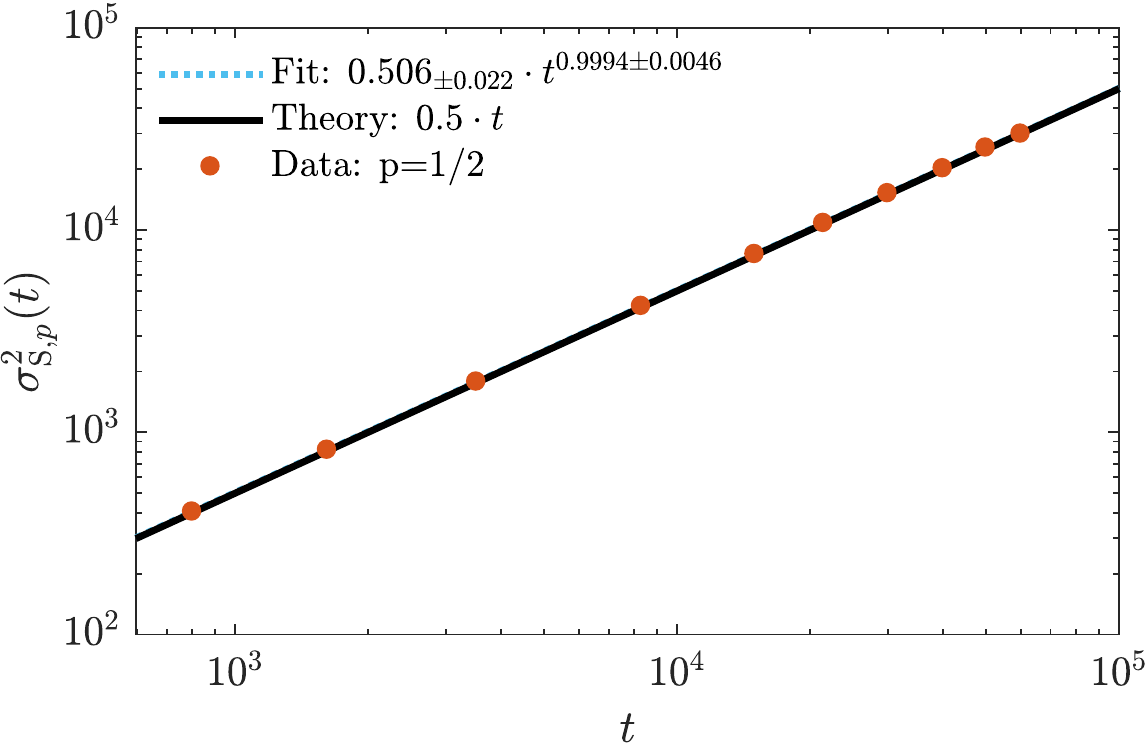}
		\caption{}
		\label{fig:1_1_sep_variance}
	\end{subfigure}
	\begin{subfigure}[t]{0.48\textwidth}
		\includegraphics[width=\textwidth]{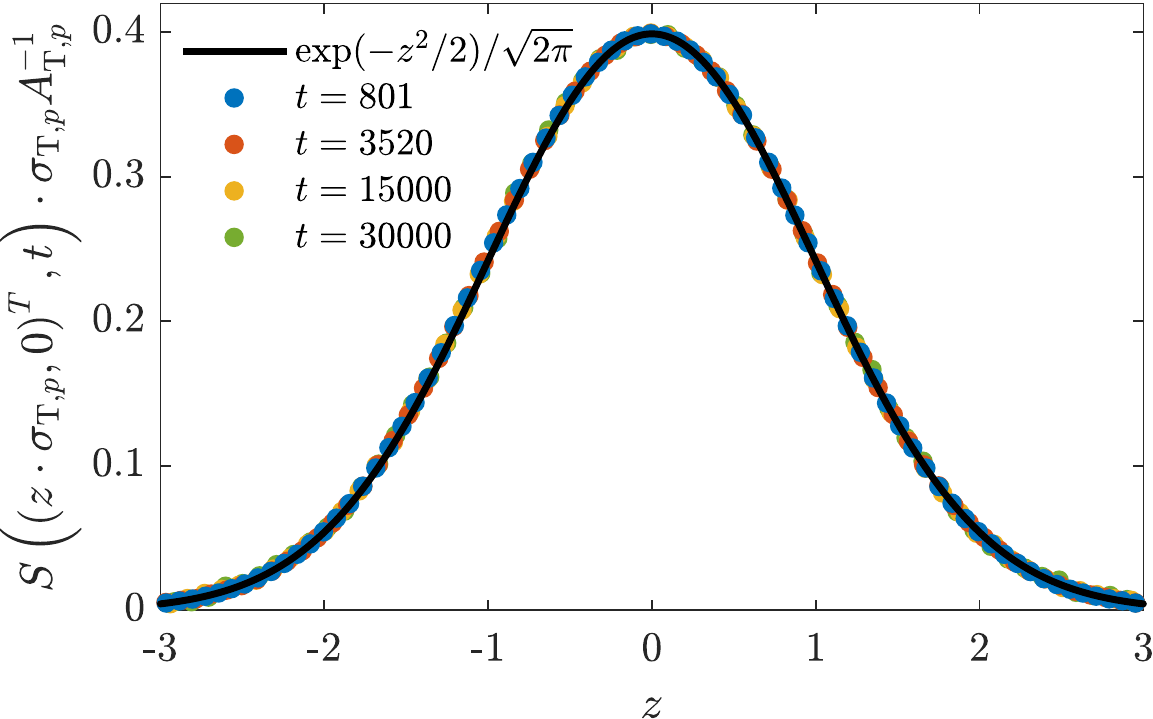}
		\caption{}
		\label{fig:1_1_tasep_collaps}
	\end{subfigure}
	\begin{subfigure}[t]{0.48\textwidth}
		\includegraphics[width=\textwidth]{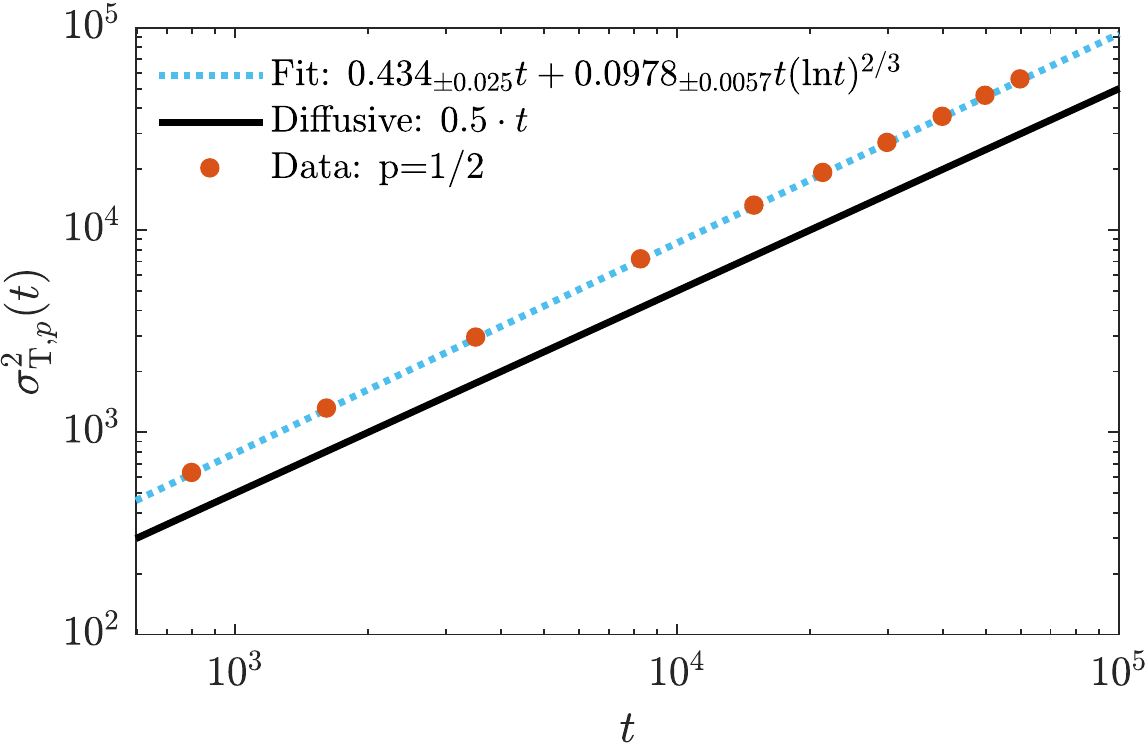}
		\caption{}
		\label{fig:1_1_tasep_variance}
	\end{subfigure}
	\caption{Scaling function analysis for the $p=\frac{1}{2}$
		system. The cross sections along the (a) $x_\SE$ and (c) $x_\TA$
		directions perfectly collapse to Gaussians. Their variances can be
		successfully fitted against the theoretical predictions: (b) shows
		clean diffusion along $x_\SE$, and (d) indicates the superdiffusive
		behavior along $x_\TA$, as the measured variance grows superlinearly. Error bars are within symbol size.}
	\label{fig:1_1_shape_analysis}
\end{figure}

The first object of analysis is the shape of the scaling
function. Before analyzing the logarithmic corrections, we want to
check that the scaling function indeed factorizes into Gaussians along
the two axes. We therefore fit the simulation data to functions of the
form
\begin{align}
\label{Eq:ScalingFunctions}
S((0,x_\SE)^T,t) = \frac{A_{\SE,\probT}(t)}{\sqrt{2 \pi
		\varS{t}{\probT}}}\exp\left[-\frac{x_\SE^2}{2
	\varS{t}{\probT}}\right], \\
S((x_\TA,0)^T,t) = \frac{A_{\TA,\probT}(t)}{\sqrt{2 \pi
		\varT{t}{\probT}}}\exp\left[-\frac{x_\TA^2}{2
	\varT{t}{\probT}}\right].
\end{align}
For the $x_\SE$ direction, we effectively have coupled
symmetric exclusion processes for which it is known that the scaling
function is a Gaussian with variance growing $\propto t$. As only a
fraction $1-\probT$ of the moves occur in this direction, we expect
$\varS{t}{\probT}=2\diffSp{\probT}t=(1-\probT)t$, which is perfectly
verified, see Fig. \ref{fig:1_1_sep_variance}. In the fit process,
$A_{\SE,\probT}(t)$ had to be included as a fit parameter, as it
contains the (so far unknown) peak height of the $x_\TA$-Gaussian. The
data in Fig. \ref{fig:1_1_sep_collaps} then perfectly collapse on the
predicted curve. 

For the fit along the $x_\TA$-axis we make use of the verified prediction
$A_{\TA,\probT}(t)=(2\pi\varS{t}{\probT})^{-\frac
	12}=(2\pi(1-\probT)t)^{-\frac 12}$ and only fit the peak width. The
data again collapse within 99\% error bars, see
Fig. \ref{fig:1_1_tasep_collaps}. From the variance against time in
Fig. \ref{fig:1_1_tasep_variance}, we can extract the TASEP diffusion
constant $\diffTp{\probT}$, for which there is no theoretical prediction, 
and the coefficient $\LogScalep{\probT}$ of the superdiffusive
logarithmic correction. In the following two subsections we will
exploit  the $\probT$-dependence of these quantities in order to extract
further information from the simulation data. 

\subsection{Scaling with the anisotropy parameter $\probT$}

\begin{figure}[h]
	\centering
	\begin{subfigure}[t]{0.48\textwidth}
		\includegraphics[width=\textwidth]{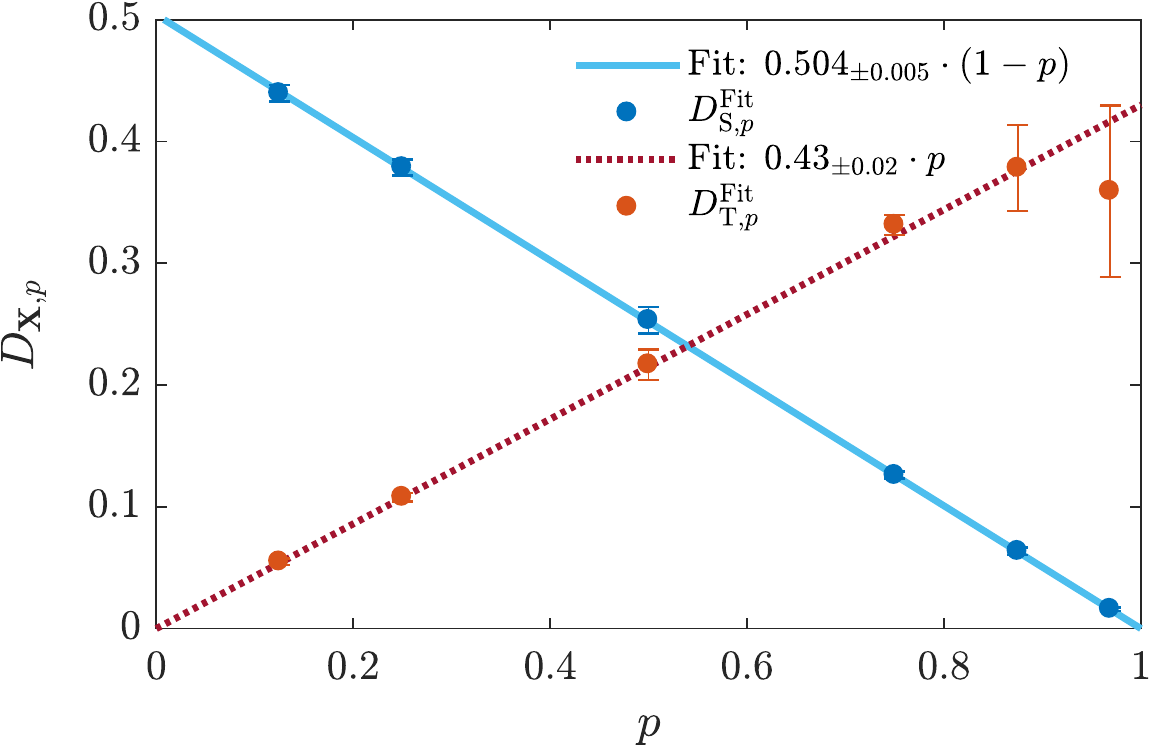}
		\caption{}
		\label{fig:p_diff_const}
	\end{subfigure}
	\begin{subfigure}[t]{0.48\textwidth}
		\includegraphics[width=\textwidth]{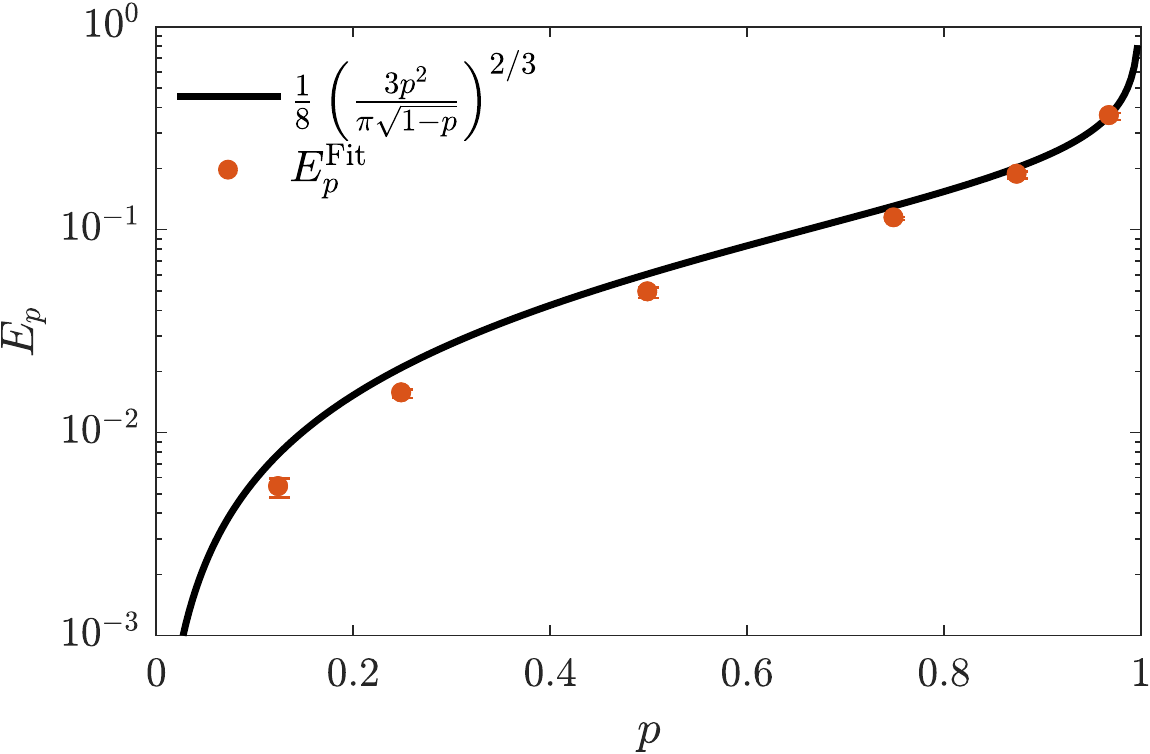}
		\caption{}
		\label{fig:p_logscale_const}
	\end{subfigure}
	\caption{(a) Diffusion constants $D_{\mathrm{X},p}$ with
          $X=\SE$ or $\TA$, and (b) the logarithmic correction
		coefficient as a function of the anisotropy parameter $\probT$,
		fitted from variance data satisfying
                $\sigma^2_{X,p}(t)\gtrsim 400$,
		resulting in lower bounds $\{3520;~1618;~801;~1618;~3520;~15000\}$ for $t$.
		This ensures that the $2\sigma$-regime of the peak spreads on at least 80
		lattice sites and the continuum approximation is justified.
		For the diffusion coefficients the expectation of a linear
		dependence on $\probT$ and $1-\probT$ is well confirmed, but the 
		logarithmic correction coefficient deviates significantly from the
		theoretical prediction. For the depicted values of $\probT$, the
		relative deviation $|E_p^{\mathrm{Fit}}-E_p|/E_p$ takes the values
		$\{0.32;~0.26;~0.19;~0.45;~0.084;~0.021\}$.}
	\label{fig:p_const_scaling}
\end{figure}

\noindent Using the fit method described in the previous subsection, one can
extract the diffusion constants $\diffTp{\probT}, \diffSp{\probT}$ 
and the coefficient of the logarithmic correction $\LogScalep{\probT}$
as a function of $\probT$. In Fig. \ref{fig:p_diff_const} we verify
the linear scaling of the SSEP and TASEP diffusion constants
$\diffSp\probT=\diffSuniv(1-\probT), \diffTp\probT=\diffTuniv\probT$
and extract the universal prefactors
\begin{equation}
\diffTuniv^{\mathrm{Fit}}=0.43\pm 0.02,~~\diffSuniv^{\mathrm{Fit}}=0.504\pm 0.005.
\end{equation} 
As advertised previously, the diffusion constant along the $x_\SE$
direction agrees with the value $\diffSuniv = \frac{1}{2}$ expected
for the SSEP. 

The overall $\probT$-dependence of the logarithmic correction coefficient
$\LogScalep\probT$ is also well described by the mode coupling
prediction (\ref{eq:sclae_parameter_log_diffision_2}), but there are significant quantitative deviations
particularly for small $\probT$
(Fig. \ref{fig:p_logscale_const}). This should not come as a surprise,
since we have seen above in Sect.~\ref{Sec:MCT} that the convergence to the
superdiffusive asymptotics is expected to be fastest for $\probT$
close to unity. Note, however, that the condition 
(\ref{eq:large_t_quantification}) is based on
the validity of continuum theory, which requires that the width of the 
correlation function is large compared to the lattice spacing along 
\textit{both} axes. This leads to a loss of accuracy when the
one-dimensional limit $p=1$ is approached too closely, as can be seen
from the estimate of $\diffTp\probT$ at $p=\frac{31}{32}$ in 
Fig.~\ref{fig:p_diff_const}. 

\subsection{Logarithmic exponent}

\begin{figure}[h]
	\centering
	\includegraphics[width=0.9\textwidth]{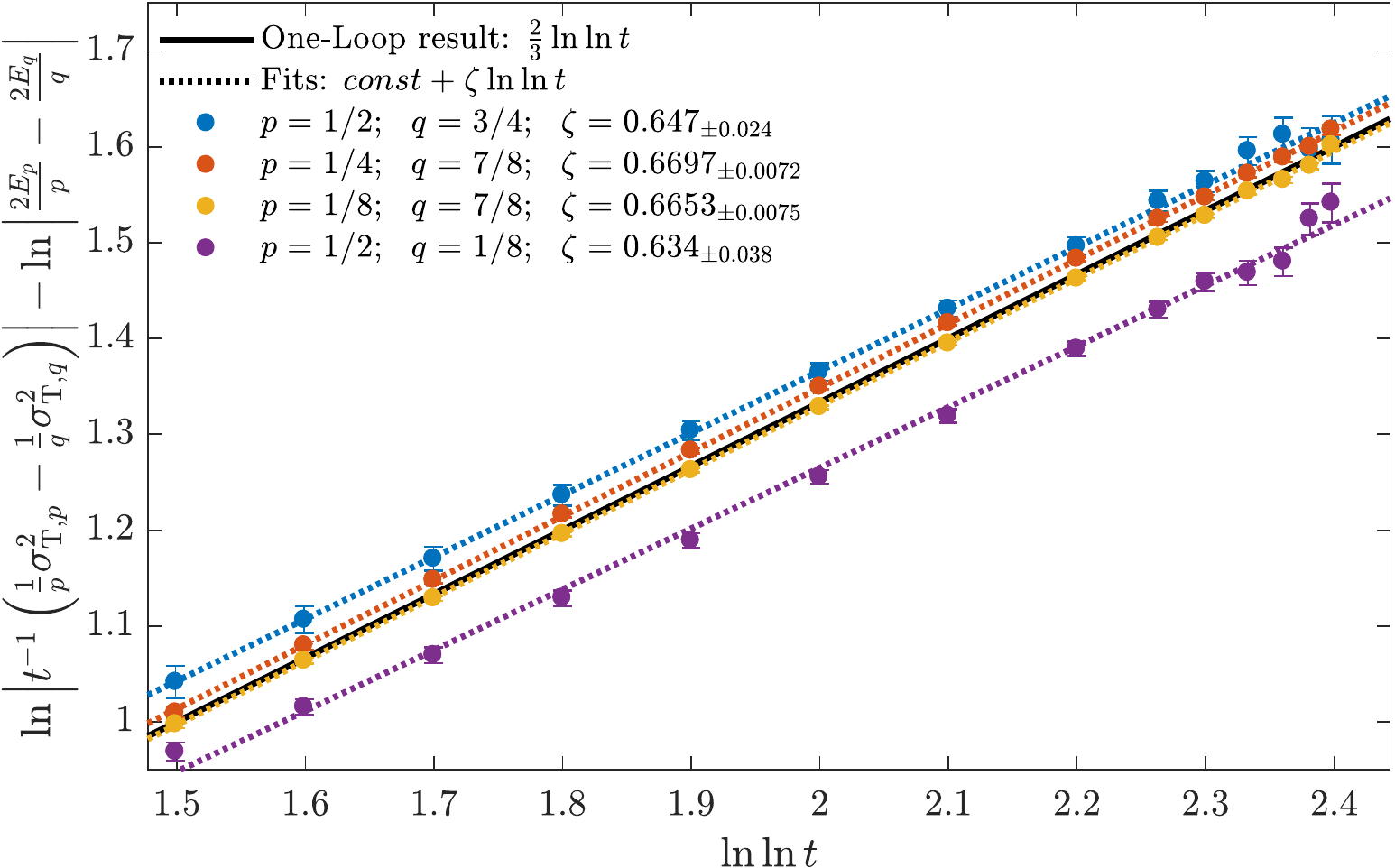}
	\caption{The logarithm of the differences between the rescaled TASEP
		variances defined in Eq.~(\ref{Eq:DiffVar}) is plotted against doubly
		logarithmic time. The slopes coincide within error with the
		theoretical prediction of $\zeta=\frac{2}{3}$. 
		Time $t$ ranges from $88$ to $60000$.}
	\label{fig:zeta_fit}
\end{figure}

So far we have determined the coefficient of the superdiffusive
correction \textit{assuming} that it has the functional form predicted by mode coupling
theory. Based on the analyses presented in the preceding subsections,
we are now prepared to extract an unbiased estimate for the logarithmic
exponent $\zeta$. 
For this, the crucial idea is to use that the diffusive part of
$\varT{t}{\probT}$ scales
linearly in $\probT$, but the logarithmic correction does
not. Therefore, for two systems with different anisotropy parameters
$\probT\neq\probTb$, the mode coupling approximation predicts
\begin{align}
\label{Eq:DiffVar}
\frac{1}{\probT}\varT{t}{\probT} - \frac{1}{\probTb}\varT{t}{\probTb} =& 2\left(\frac{\LogScalep{\probT}}{\probT}-\frac{\LogScalep{\probTb}}{\probTb}\right) t (\ln t)^{\zeta} \nonumber \\
=& 2\LogScaleuniv
\left(\left(\frac{\probT}{1-\probT}\right)^{\frac{1}{3}}-\left(\frac{\probTb}{1-\probTb}\right)^{\frac{1}{3}}\right)
t (\ln t)^{\zeta},
\end{align}
which allows to fit $\zeta$ reliably from the data. As shown in
Fig. \ref{fig:zeta_fit}, within error bars this analysis fully
confirms the predicted exponent $\zeta=\frac{2}{3}$. By contrast, 
the prefactor of the logarithmic power law displays some deviations
from the prediction of mode coupling theory. However, since these
deviations correlate with the errors in the exponent
$\zeta$, they might be attributable to finite time corrections.   

% #########################################################
% ####################	DISCUSSION		###################
% #########################################################
\section{Discussion}
\label{Sec:Discussion}

The work of BKS \cite{Beij85} stands at the beginning of a spectacular development that has revealed a universe of interrelated one-dimensional systems,  all of which are characterized by the 
superdiffusive spreading of fluctuations as $t^{2/3}$ \cite{Corwin2012,KPZ1986,KriKru2010,GSS2017}. In comparison, the ``other $\frac{2}{3}$ -law'' discovered by BKS, which is 
associated with the value of the logarithmic correction exponent $\zeta$ in two-dimensional driven diffusive systems, has received much less attention. 
Here we have shown that a precise numerical verification of this subtle phenomenon is now possible.

The newly found estimator \eqref{single-estimator} based on the
concept of coupling gave us an exciting new tool to push the limits of
numerical exploration, such that logarithmic corrections could be made
visible. In accordance with the computations of \cite{Beij85}, we find
strong numerical evidence that (i) the shape of the scaling function
is Gaussian, (ii) diffusion is normal along the transverse $x_\SE$-axis, and
(iii) the logarithmic correction term of the variance along the
direction $x_\TA$ of the drive is $\propto t (\ln t)^{\frac23}$.
The only significant deviations from the theoretical predictions were
found in the value of the logarithmic correction coefficient
$\LogScale$ for small $\probT$. Since the overall dependence
on $\probT$ is in good agreement with the predictions, these
deviations may simply reflect the slow convergence to the asymptotic
behavior that is expected for small $\probT$.

In future work it would be of obvious interest to go beyond the
particular version of the ASEP considered here in order to establish the
broader universality of the logarithmic superdiffusivity. Whereas
certain generalizations of the model can be implemented
straightforwardly, others would require further nontrivial innovations
in the numerical algorithm. For example, the present method could be
easily extended to the case where particles move strictly
asymmetrically along the two lattice axes with probability $q$ and
$1-q$, respectively, such that the direction of the drive varies
continuously with $q$. On the other hand, in the absence of  
the particle-hole symmetry imposed by the condition
$\density=\frac12$, numerical analysis is much harder, because 
the drift of density fluctuations will strongly increase the
systematic errors \cite{GSS2017}. In this sense the limitations of the numerical
approach are somewhat similar to those encountered in the rigorous
analysis of the problem \cite{Yau04}, which is also restricted to $\rho =
\frac12$.  

Nevertheless, the idea of using coupling to evolve large ensembles of
systems in parallel might offer great opportunities for applications
to other problems. 
In our simple single species model, the coupling between different
realizations is complete in the sense that second class particles
cannot be generated at all, and our methodology is clearly most
efficient for models that share this property. Other applications
might be more complex, in that creation of second class particles is
allowed. Estimators making use of second class particle extinction due
to coupling then still offer a stronger decline in variance until the 
extinction and creation rate reach equilibrium. More generally, we
hope that our work will stimulate further investigations into the
dynamics and applications of second class particles, including a theoretical
explanation of the power law decay reported in Eq.~(\ref{Eq:DensityDecay}). 

\paragraph{Acknowledgments.} We thank Herbert Spohn for his inspiration and leadership, and wish him many years of enjoyable and productive research.
This work was supported by Deutsche Forschungsgemeinschaft (DFG) under
grant SCHA 636/8-2 and by the University of Cologne through the UoC
Forum \textit{Classical and quantum dynamics of interacting particle
  systems}. 
% \appendix

\end{document}